\documentclass[useAMS,usenatbib,usegraphicx]{mn2e}

\title[A Link Between the Semi-Major Axis and Stellar Metallicity]
{A Link Between the Semi-Major Axis of Extrasolar Gas Giant Planets and
Stellar Metallicity}
\author[R. Pinotti et al.]{
R. Pinotti,$^{1,2}$ \thanks{E-mail: rafaelpinotti@petrobras.com.br}
L. Arany-Prado,$^{1}$ \thanks{E-mail: lilia@ov.ufrj.br}
W. Lyra$^{1,3}$
and G. F. Porto de Mello$^{1}$\\
$^{1}$ Observat\'orio do Valongo, Universidade Federal do Rio de Janeiro,
       Lad. Pedro Ant\^onio 43, RJ, 20080-090, Brazil\\
$^{2}$ PETROBRAS, REDUC/OT, Rod. Washington Luiz km 113,7, Duque de Caxias, RJ,
       25225-010, Brazil\\
$^{3}$ Department of Astronomy \& Space Physics, Uppsala Astronomical Observatory, Box 515, 751 20, Sweden\\
}
\begin{document}

\date{Accepted 2005 August. Received 2005 July 14; in original form 2005 January 14}

\pagerange{\pageref{firstpage}--\pageref{lastpage}} \pubyear{2005}

\maketitle

\label{firstpage}

\begin{abstract}
The fact that most extrasolar planets found to date are orbiting
metal-rich stars lends credence to the core accretion mechanism of gas giant
planet formation over its competitor, the disc instability mechanism.
However, the core accretion mechanism is not refined to the point of
explaining orbital parameters such as their unexpected semi-major axes and
eccentricities. We propose a model, which correlates the metallicity of the
host star with the original semi-major axis of its most massive planet, prior
to migration, considering that the core accretion scenario governs giant gas
planet formation. The model predicts that the optimum regions for planetary
formation shift inward as stellar metallicity decreases, providing an
explanation for the observed absence of long period planets in metal-poor
stars. We compare our predictions with the available data on extrasolar
planets for stars with masses similar to the mass of the Sun. A fitting procedure 
produces an estimate of what we define as the Zero Age
Planetary Orbit (ZAPO) curve as a function of the metallicity of the star. 
The model also hints that the lack of planets circling metal-poor stars 
may be partly caused by an enhanced destruction probability during the 
migration process, since the planets lie initially closer to the central stars.
\end{abstract}


\section{Introduction}

The discovery of the first extrasolar planet around 51 Pegasi (Mayor \& Queloz 1995) prompted an intensive search effort that continues today. Currently a hundred and sixty of them are catalogued, though the definition of planet may not be suited to some, whose masses place them more conveniently in the brown dwarf realm. The ultimate goal, the detection of a planet like our own, lies in a foreseeable future, as promised by plans of sophisticated space-based instruments capable of detecting Earth mass planets.

Even with a sample of gas giant planets at hand we face more than a few challenges regarding the main competing theories of gas giant planet formation, namely, core accretion and disc instability. None of them predicted the existence of gas giants so close to their parent stars, or the pronounced eccentricity of some of their orbits. There is, however, a feature of that sample that favours the core accretion mechanism: most of the planets are circling metal-rich stars, and this relation has been dismissed as a possible outcome of selection effects, or the result of pollution by the absorption of protoplanetary disc material and/or planets and/or disc leftovers such as planetesimals and comets (Santos et al. 2003; Pinsonneault, DePoy \& Coffee 2001). Santos, Israelian \& Mayor (2004) further argue that the frequency of planetary companions is remarkably constant ($\sim\,$3\%) up to solar metallicity, rising very rapidly afterwards and reaching $\sim\,$30\% for twice the solar metallicity, a result that suggests the existence of threshold metallicity values for the efficiency of planet building. 

Even studies that indicate some influence of selection effects and pollution (Murray \& Chaboyer 2002; Laughlin 2000; Israelian et al. 2001) concede that the high metallicity is at least partly caused by the primordial metal-rich nebula that formed the stars. The fact that the photometric survey (Gilliland et al. 2000) of $\sim\,$34,000 stars in the globular cluster 47 Tucanae ([Fe/H] =$\,-$0.7) has found no transiting short-period planets is also an important indicator of the link between stellar metallicity and gas giant planet formation. The researchers expected to find $\sim\,$17 planets, based on the statistics of radial velocity surveys. However, the result of this survey should not be easily taken as typical for low metallicity stars because the very nature of the cluster may have created a bias. The high star density may have imposed extra restriction regarding planet formation and stability, since the radiation from the expected number of O stars can be sufficient to remove circumstellar discs before planets could form (Armitage 2000). Moreover, close encounters between stars in the early stages of the globular cluster can also disrupt a disc (Bonnell et al. 2001), or eject the planets that were allowed to form (e.g., Sigurdsson 1992), leading to a population of free-floating planets that may or may not remain bound to the cluster, depending on the velocity of the ejection (e.g., Hurley \& Shara 2002) . Altogether, these processes might explain the lack of planets in 47 Tuc. It is worth noting, on the other hand, that another recent photometric survey of 47 Tuc (Weldrake et
al. 2005), more sensitive to the uncrowded outer regions, indicates that system metallicity is the dominant effect.

Since both the accretion disc and the parent star are products of the same gas and dust cloud, a relationship between primordial metallicity and the frequency of planet finding would be a natural outcome in the core accretion scenario (Pollack et al. 1996), which holds that a rocky nucleus of around 10 Earth masses is formed before the massive infall of gas completes the formation of a gas giant. The disc instability model, on the other hand, is based on a self-gravitating mass of gas that is formed in the rotating accretion disc and is fairly insensitive to metallicity (Boss 2002a). One of the main merits of the disc instability model is the short formation time ($10^2$ to $10^3$ years) of protoplanets, observed in the simulations by Boss (2002b) and Mayer et al. (2002), compared with the considerably longer time in a core accretion scenario (up to 10$\,$Myr), which would in principle violate the age constraint for accretion discs, based on observation of young stellar objects (Haisch, Lada \& Lada 2001; Bary, Weintraub \& Kastner 2003). However, the simulations do not account for the long-term survival of the protoplanets, or the origin of the instability. Moreover, simulations incorporating stochastic core migration through the protoplanetary disc (Rice \& Armitage 2003) show that a Jupiter mass planet can be formed within 10$\,$Myr.

If the core accretion mechanism is indeed the one that regulates the formation of gas giant planets, as well as of rocky ones, then the existence of gas giant planets around metal-rich stars is probably but one consequence.

\section[]{Optimum environment for planet formation}

Let us consider the key radial variables for planet formation in the accretion disc, disc temperature and surface density. The disc temperature drops continuously as the distance to the star increases, following a power law. For example, a crude profile can be derived from gravitational potential energy conservation and blackbody radiation (Hartmann 1998), showing that the disc temperature is proportional to $r^{-3/4}$, where $r$ represents the radius; a similar relation can be written considering only the radiation from the central star that is intercepted by the disc.

Since the gravitational pull of the embryo gas giant planet gets more efficient for low gas temperature, we should expect gas giants to form at considerable distances from their host stars. On the other hand, the roughly spherical form of the contracting nebula dictates that the disc surface density has a decreasing radial profile (Hartmann 1998; Pringle 1981), so that at large distances from the star the advantage of low temperatures is hampered by the lack of available matter. The ``optimum formation radius" is therefore a compromise between these two parameters. We should then expect that at this radius the probability of planet formation is maximum, or, considering multiple planet systems, that at this point the most massive planet is expected to be formed.

This scenario is consistent with an embryo planet already pulling gas and dust, but the initial step of planet formation, according to the core accretion model, would involve not the surface density, but more precisely the {\em dust} surface density, since a rocky core is formed in the first place. The dust surface density also decreases as distance increases, so that we have again an ``optimum formation radius".

But here we hit a possible connection between the stellar metallicity and the semi-major axis of our embryo planet; for we could then expect that for stars with higher metallicity (and considering the stellar and accretion disc masses equal), the radial profile of the dust surface density is altered.

According to the studies of the radial flow of dust particles in accretion discs, by Takeuchi \& Lin (2002), medium size particles (10 to 100 $\mu$m) will migrate inward rapidly (orbital decay time $\sim\,$$10^6\,$yr for 100$\,\mu$m particles), distorting the radial profile in a way that in the inner part of the disc (up to $\sim\,$15-20$\,$AU), which is the relevant distance range for the formation of the class of planets we are discussing, it will get flatter (see figure$\,$9a of Takeuchi \& Lin 2002), becoming steeper further out. Larger particles ($\ge\,$1$\,$mm) will concentrate in an even narrower region (upper bound $\sim\,$1$\,$AU), beyond which the dust surface density falls to a value around one order of magnitude below that of 100 $\mu$m particles.

If we also take into account that gas giants are not believed to form through core accretion beyond $\sim\,$20$\,$AU (Mayer et al. 2002), since the formation of a solid core would take too long with the surface densities involved, the profile in the region of potential gas giant planet formation is flatter compared with that of the outer region, and we can infer that it will get even flatter in a higher metallicity disc, since more dust particles will migrate radially inward. This phenomenon will push the ``optimum formation radius" outward, since the competing effect of a lower temperature will tend to get the upper hand.

Thus, considering the physical variables of both the accretion disc and the parent star that were taken into account, there may be a significant correlation, in a sufficiently large sample of planetary systems, between the initial semi-major axis of the most massive gas giant planet in a given planetary system and the metallicity of its parent star, since the most massive one would probably be formed at or near the optimum radius. Such behaviour would be constrained at great distances, where the dust surface density becomes too low, and at very low radii, where the high temperature prevents the massive accretion of gas from taking place, leaving only small, rocky planets like the Earth. The new planets circling very metal-rich stars and very metal-poor ones would then tend to be located at two asymptotic semi-major axis (SMA) values. These arguments lead naturally to the suggestion that a growth curve might exist.

This model can be parametrized mathematically by the following reasoning.

Let $P$ be the probability of giant gas planet formation as a function of radius; it can be assumed as being proportional to the dust surface density, $\sigma_S$ (Lineweaver 2001; Wetherill 1996), since the planet formation process requires the {\em a priori} agglomeration of a rocky planetary embryo, whose rate of growth is directly proportional to the surface mass density of solids (Pollack et al. 1996; De Pater \& Lissauer 2001), which in turn is directly related to the dust surface density at the time of the formation of solid planetesimals. Although the results by Pollack et al. 1996 indicate such proportionality only
during the so called Phase I, which is short compared with Phase II, we understand that $P$ depends first of all on the formation of a rocky core. Moreover, the time spent in Phase I can be substantially increased
depending on the initial conditions, as shown by Pollack et al. in the case of the formation of Uranus. $P$ can also be assumed as being inversely proportional to the midplane disc temperature, $T$, since the gas accretion rate that ultimately forms the gas giant is directly proportional to the gas density in the immediate
vicinity of the protoplanet (Pollack et al. 1996). The gas density, in turn, can be considered proportional to $1/T$, as dictated by the ideal gas law, in a nebula with gas in pressure equilibrium, at least until the beginning of the runaway gas accretion. If the gas accretion rate is too small, the formation time exceeds the lifespan of the gas in the protoplanetary disc and no gas giant will be formed. In the simulation conducted by Pollack et al. (1996), the temperature also influences the gas accretion rate in a more subtle way, as explained in the beginning of this section: the effective radius of the planet is defined by the boundary beyond which the thermal energy of the gas exceeds the gravitational binding energy, and the gas accretion rate is directly proportional to the effective radius, so a planet travelling through a cooler gas will tend to grow faster. Here we note that the statement by Pollack et al. (1996) that their results are insensitive to $T$ and gas density is not valid in the context of our approach, since they use pairs of $T$ and gas density for the same protoplanetary nebula (Mizuno 1980, Bodenheimer \& Pollack 1986), in which both variables decrease with radius, whereas we consider that, for a given radius, a cooler nebula will correspond to an increase in gas density.

Thus, we can write
\begin{equation}
   P \propto (\sigma_S/T).
   \label{eq1}
\end{equation}

Considering that the radial profile of $\sigma_S$ (De Pater \& Lissauer 2001) is similar to that of surface density as adopted in simulations by Takeuchi \& Lin (2002) and Boss (2002b), then
\begin{equation}
   \sigma_S \propto r^{-\alpha},
   \label{eq2}
\end{equation}
  where $\alpha$ varies between approximately 0.5 and 1.5.

The disc temperature profile is also assumed to follow a power law (Mayer et al. 2002)
\begin{equation}
   T \propto r^{-\beta} + t,
   \label{eq3}
\end{equation}
where in Boss (2002b) $\beta\sim 2$ and $t$ is a constant (at least for a given stellar mass). However, in the more recent simulation by Nomura \& Millar (2005) $\beta\sim 1$, which is in good agreement with the earlier simulations of D'Alessio et al. (2001). In Eq.$\,$(\ref{eq2}) and (\ref{eq3}) $r$ is given in arbitrary units. Note that even if we consider that the dependence of $P$ on the temperature follows a power law, this can be incorporated into the parameter $\beta$. 

In order to take into account our hypothesis that the probability is also a function of the metallicity, that is $P=P(r,Z)$, where $Z\,\equiv\,{\rm [Fe/H]}$, we assume that $\alpha=\alpha(Z)$ and $\beta=\beta(Z)$, in such a way that the values mentioned above (for $\alpha$ and $\beta$) belong to the high metallicity range of the functions we are searching for. An increase in $Z$ should cause an increase in the dust surface density, lowering the value of $\alpha$, and also an increase in the disc opacity, which lowers the temperature gradient. Asymptotic values for $\alpha$ and $\beta$ should be assumed when $Z$ is large, since the boundary conditions ruling the heat exchange process (central star photosphere
temperature, disc mass of finite extent, and interstellar matter temperature) always require decreasing radial profiles for $T$ and $\sigma_S$. On the other hand, large negative values of $Z$ must result in large values for $\alpha$ and $\beta$.

The optimum value of (1) is calculated by setting 
\noindent \begin{displaymath}
{\rm d}P = {\partial P\over\partial r}\ {\rm d}r +
              {\partial P\over\partial Z}\ {\rm d}Z = 0, \nonumber
\end{displaymath}
so, ${\partial P/\partial r}=0$ and ${\partial P/\partial Z}=0$.
Using Eq.$\,$(\ref{eq2}) and (\ref{eq3}), from ${\partial P/\partial r}=0$
we obtain the optimum formation radius
\begin{equation}
   r^\beta_{\rm opt} = \left({\beta-\alpha\over t\ \alpha}\right).
   \label{eq4}
\end{equation}
As expected, a flatter surface density profile will push the optimum radius outward. Using $\partial P/\partial Z=0$ the same procedure gives
\begin{equation}
   r^\beta_{\rm opt} = \left({{\rm d}\beta-{\rm d}\alpha\over t\ {\rm d}\alpha}\right),
   \label{eq5}
\end{equation}
so, using Eq.$\,$(\ref{eq4}) and (\ref{eq5}), we arrive at $\beta\ {\rm d}\alpha = \alpha\ {\rm d}\beta$, which is valid for a constant ratio between $\alpha$ and $\beta$, that is, $\beta/\alpha = \beta_i/\alpha_i$, where $\beta_i$ and $\alpha_i$ are, respectively, any inferior limit
of the sum over some range of the functions $\alpha(Z)$ and $\beta(Z)$. It is worth noting that even if we consider only ${\partial P/\partial r}=0$ (and not necessarily $\partial P/\partial Z=0$) and once it is supposed that a successful planetary formation process is the one with simultaneous optimum or near-optimum values of $\alpha$ and $\beta$, that is
\noindent \begin{displaymath}
   {\beta-\alpha\over t\ \alpha} \approx {(\beta + {\rm d}\beta)-(\alpha + {\rm d}\alpha)\over t\ (\alpha +{\rm d}\alpha)}\ ,
\end{displaymath}
we still obtain $\beta\ {\rm d}\alpha \approx \alpha\ {\rm d}\beta$.

The relation between $\alpha$ and $\beta$ can be generalized as
\begin{equation}
   {{\rm d}\alpha\over \alpha} = {{\rm d}\beta\over \beta} = -\,c\,f(Z)\ {\rm d}Z,
   \label{eq6}
\end{equation}
where $c$ is a positive constant and $f$ is some function of the primordial metallicity (which is mirrored by the metallicity of the star prior to potential pollution effects). We have put the negative sign because of our basic assumption that $\alpha$ (or $\beta$) increases as $Z$ decreases. A general form for the so far unknown $f$ in Eq.$\,$(\ref{eq6}) which takes into account the constraints mentioned before and meets the mathematical requirements is
\begin{equation}
   f(Z) = \left(1+e^{\zeta(Z)}\right)^{-1}, 
   \label{eq7}
\end{equation}
where
\begin{equation}
 \zeta(Z) \equiv  c\,(Z-Z_0)\,  .
 \label{eq7b}
\end{equation}
The substitution of $Z$ by $(Z-Z_0)$ accounts for the fact that the value $\zeta=0$ does not correspond necessarily to the solar metallicity. Note that $f$ varies from 1 to 0 as $Z$ varies respectively from $-\infty$ to $+\infty$. Then Eq.$\,$(\ref{eq6}) states that for large values of $Z$, $\alpha$ and $\beta$ have asymptotic constant values and for low values of $Z$ both parameters and their rate of change, d$\alpha$ and d$\beta$, go to $+\infty$.

Integrating Eq.$\,$(\ref{eq6}) using (\ref{eq7}) and (\ref{eq7b}) we have
\noindent \begin{displaymath}
   \int^\infty_\alpha {\rm d}\ln \alpha' = \int^\infty_\beta {\rm d}\ln \beta' =
   \int^{\zeta}_\infty {(1+e^{\zeta'})^{-1}\,{\rm d}\zeta'}\  ,
\end{displaymath}
which gives the functions we have been looking for, respectively
\begin{equation}
  \label{eq8}
\begin{array}{ll}
  \alpha(Z)  = \alpha_{\rm a}\left(1+e^{-\zeta(Z)}\right) , \\
  \beta(Z)  = \beta_{\rm a}\left(1+e^{-\zeta(Z)}\right) ,
\end{array}
\end{equation}
were $\alpha_{\rm a}$ and $\beta_{\rm a}$ are the asymptotic values at high metallicity and $\zeta$ is given by Eq.$\,$(\ref{eq7b}).

Substituting Eq.$\,$(\ref{eq8}) in (\ref{eq4}) we have
\begin{equation}
   r_{\rm opt}(Z) = \left({\beta_{\rm a}-\alpha_{\rm a}\over t\ \alpha_{\rm a}}\right)^{1/\beta\mathstrut(Z\mathstrut)} \,\,\,\,\,\, ({\rm in\,\, arbitrary\,\, units}).
   \label{eq9}
\end{equation}
Thus, the optimum radius (which can be replaced by the SMA of unmigrated planetary orbits) is then linked to the metallicity $Z$. The population of young planets not yet influenced by migration will tend to follow a curve, dubbed ZAPO, for zero age planetary orbit, in a metallicity versus SMA diagram. 

From Eq.$\,$(\ref{eq9}), the minimum and maximum values of $r_{\rm opt}$ are, respectively
\begin{eqnarray}
\label{eq10}
  \lim_{Z\to -\infty} r_{\rm opt}(Z)&=&1 , {\rm \ since\ }\beta\to\infty\ , \nonumber 
\\
\lim_{Z\to +\infty} r_{\rm opt}(Z) &=& \left({\beta_{\rm a}\mathstrut-\alpha_{\rm a}\mathstrut\over t\ \alpha_{\rm a}}\right)^{1/\beta_{\rm a}}.
\end{eqnarray}

To rescale 1 arbitrary unit (in the above equation) to AU we take 
\begin{eqnarray}
\label{eq11}
  1\,  {\rm arbitrary\,  unit} = \gamma^{-1}\, {\rm AU} = \gamma^{-1}\,  214.94\, {\rm R_\odot} , 
\end{eqnarray}
so that $\gamma$ fixes the most probable minimum distance $r_{\rm opt}$ where a massive planet can be formed at a very low metallicitiy. Using Eq.$\,$(\ref{eq9}) and (\ref{eq11}), Eq.$\,$(\ref{eq3}) can be written as
\begin{eqnarray}
\label{eq12}
   T \propto  \left( {r\over\gamma}\right)^{-\beta(Z)} + {\beta_a-\alpha_a\over \alpha_a}\,\,\left({r_{\rm opt}\over\gamma}\right)^{-\beta(Z)}   \,\, ,
\end{eqnarray}
where $(r/\gamma)$ is given in AU. One remark on self-consistency is warranted here. The second term of Eq.$\,$(\ref{eq12}) is equivalent to $t$ in Eq.$\,$(\ref{eq3}), supposed constant. This second term is actually very slightly metallicity dependent, over the metallicity interval in which $P(r)$ is non-negligible. As will be shown below, in the discussion of the fitting process, $P(r)$ vanishes rapidly for [Fe/H]$\,\lower 3pt\hbox{$\buildrel<\over\sim$}-1$, and therefore this second term can be considered as constant over the metallicity interval of the data. As we describe on the next section, although the introduction of the scale factor does not significantly alter the possible solutions of Eq.$\,$(\ref{eq9}), the slight dependence on $Z$ in Eq.$\,$(\ref{eq12}) results in temperature profiles for $Z$ down to [Fe/H]$\,\,\cong-1$ which are more physically meaningful than that given by Eq.$\,$(\ref{eq3}).

In order to draw to a crude estimate of a migration process which would displace the entire ZAPO curve to a lower value and form different populations of migrated planets, we introduce in Eq.$\,$(\ref{eq9}) the fraction $n$, so that 
\begin{eqnarray}
\label{eq13}
  r(Z,n)=n\,r_{\rm opt}(Z)\, .
\end{eqnarray}
Here we are considering the so-called type II migration mechanism, which dictates that the planet(s) formed is (are) large enough to open a gap in the protoplanetary disc, moving radially in lockstep with the gaseous disc. This migration process would eventually stop (Matsuyama et al. 2003), and leave the planet(s) in a smaller, but safe, orbit.

This simple mathematical development, based on optimal conditions, is atemporal by nature, and can not be derived from studies of the evolution of planetesimal surface density profiles (e.g. Youdin \& Shu 2002), because they do not provide answers as to when and where the rocky cores form. As for the physical credibility of Eq.$\,$(\ref{eq8}), one would think that when comparing the surface density $\alpha(Z)$ with the temperature profile $\beta (Z)$, the latter would tend to have a much weaker dependence on the migration of dust particles than the surface density. In fact, that $\alpha(Z)$ and $\beta (Z)$ show the same functional relation should be seen only as a rough approximation of physical reality. For example, for very low values of $Z$, $\beta (Z)$ would become very large, and the disc would become radially isothermal, which does not correspond to reality. However, even if we do not consider that $\beta (Z)$ is optimized in $Z$, and choose an arbitrary smooth function that varies with $Z$ between two finite asymptotic values, and substitute it in Eq.$\,$(\ref{eq4}), we still obtain a curve very similar to the one obtained by Eq.$\,$(\ref{eq9}), only that new adjustment parameters for the model will appear. Therefore, for the purposes of this work, the rather simplistic form of $\beta (Z)$ and $\alpha(Z)$ is robust enough so that our main conclusions are not affected.

\section{Results and discussion}

We now proceed to fit the model to the observed distribution of planetary SMA as a function of stellar metallicity. Since we are dealing with the optimum semi-major axis for planet formation, we expect that the most massive planet (in each planetary system) will tend to be formed at that locus, and the Doppler shift technique can discover the most massive ones orbiting a given star. There is of course the possibility that massive planets stay hidden in long orbits, beyond the sensitivity of the technique, or because the time line available is not currently long enough to enable detection. This is unlikely, since planets with periods reaching $\sim\,$11$\,$yr have been uncovered. Their apparent absence in stars more metal-poor than [Fe/H]$\,\lower 3pt\hbox{$\buildrel<\over\sim$}-0.4$ (Santos et al. 2003), however, is noteworthy. Against the possible explanation of this absence as arising from the larger radial velocity uncertainties in metal-poor stars due to their shallower spectral lines, Santos et al. (2003) show that the mean uncertainty in the radial velocity determinations of the CORALIE spectrograph is only 1-2 m/s higher for metal-poor stars, as compared with metal-rich ones.

\begin{figure}
 \includegraphics[]{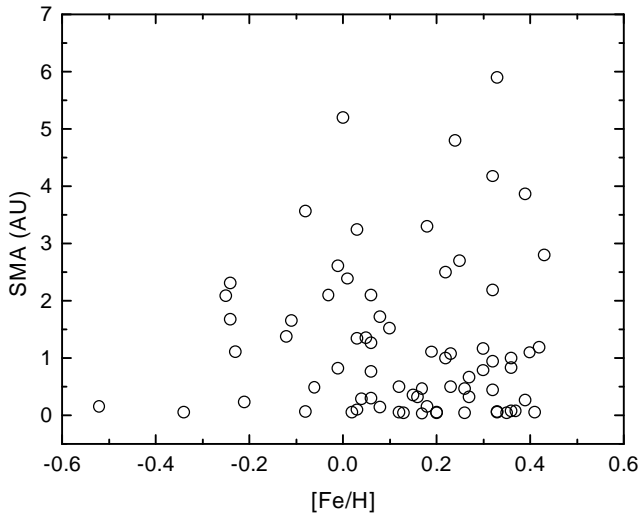}
 \caption{Stellar Metallicity versus Semi-Major Axis (SMA) for extrasolar planets orbiting stars with masses inside the range of $1\pm 0.2$ solar mass. In the case of multiple systems, only the most massive planet was chosen.}
\end{figure}
\begin{figure}
 \includegraphics[]{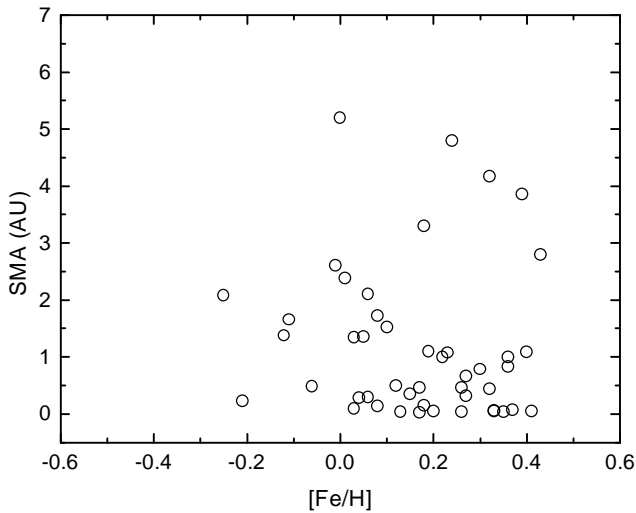}
 \caption{The same as Fig.$\,$1, taking the range of $1\pm 0.1$.}
\end{figure}

A possible source of selection effect is the fact that very short period orbits are not occupied by massive planets (Zucker \& Mazeh 2002, Santos et al. 2003), which may indicate that the migration process is more efficient for less massive planets. However, this phenomenon does not affect our main conclusions, as will be seen below. 

In our analysis we consider, for each star that harbours more than one planet, only the most massive one and assume that there is no significant probability of finding still more massive planets. On the other hand, we disregard  the ``planets" with minimum mass above 10 Jupiter masses (${\rm M}_{\rm J}$), since this is likely to be the upper limit for the mass of a planet (Jorissen, Mayor \& Udry 2001). Besides, if we consider the critical mass to be 13$\,{\rm M}_{\rm J}$, and an aleatory distribution of orbit inclinations, the mean value of the true mass is around 1.3 times the minimum mass, and we have again a minimum mass of 10$\,{\rm M}_{\rm J}$ as a constraint.

Another point to be considered is the mass range of parent stars, from 0.32 to 1.60 solar masses. Our model does not depend on the stellar mass, although it influences considerably the temperature profile of the protoplanetary disc, not to mention its mass surface density. Therefore, we separate the planet harbouring stars according to mass ranges, in order to filter out this influence. Since the search for extrasolar planets is directed mainly at stars similar to the Sun, we limit the analysis to stars whose masses fall within two ranges of the solar mass, $\pm 20\%$ (72 stars, Fig.$\,$1) and $\pm 10\%$ (46 stars, Fig.$\,$2). The former accommodates reasonably well the mass determination uncertainties, which are of the order of $\sim\,$0.05 solar masses. As we show in the following paragraphs, our conclusions are not affected by the choice of this mass interval.

Data on orbital parameters was extracted from the Extrasolar Planets Encyclopedia, maintained by Jean Schneider (Schneider 2005); the metallicity values were taken from the literature and from the Extrasolar Planets Encyclopedia. Table I contains all data used in this work and the references for the data used.

\begin{table*}
 \centering
 \begin{minipage}{160mm}
  \caption{Stellar and planetary data}
  \begin{tabular}{lcrccclcrccc}
\hline\hline
Star & Star & [Fe/H] & Ref. & Planet & Planet     & Star & Star & [Fe/H] & Ref. & Planet & Planet     \\
     & Mass &        &      & SMA    & Min. Mass  &      & Mass &        &      & SMA    & Min. Mass  \\
     & ${\rm M}_\odot$ &   &      & (au)   & ${\rm M}_{\rm J}$      &      & ${\rm M}_\odot$ &   &      & (au)   & ${\rm M}_{\rm J}$      \\
\hline
Gl 876      & 0.32 &   +0.00 & 4 & 0.21  & 1.98 & HD 73526   & 1.05 &   +0.27 & 1 & 0.66 & 3.0   \\
HD 114386   & 0.54 & $-$0.08 & 1 & 1.62  & 0.99 & HD 117618  & 1.05 &   +0.04 & 4 & 0.28 & 0.22  \\
HD 128311   & 0.61 &   +0.03 & 1 & 1.74  & 3.24 & HD 137759  & 1.05 &   +0.03 & 4 & 1.34 & 8.64  \\
HD 192263   & 0.69 & $-$0.02 & 1 & 0.15  & 0.72 & HD 217014  & 1.05 &   +0.20 & 1 & 0.05 & 0.46  \\
HD 13445    & 0.70 & $-$0.24 & 1 & 0.11  & 4.00 & HD 195019A & 1.06 &   +0.08 & 1 & 0.14 & 3.43  \\
HD 41004A   & 0.70 & $-$0.09 & 4 & 1.31  & 2.3  & HD 4203    & 1.06 &   +0.40 & 1 & 1.09 & 1.64  \\
HD 22049    & 0.73 & $-$0.13 & 1 & 3.3   & 0.86 & HD 95128   & 1.07 &   +0.06 & 1 & 2.1  & 2.41  \\
HD 37124    & 0.75 & $-$0.38 & 1 & 2.5   & 1.2  & HD 134987  & 1.08 &   +0.30 & 1 & 0.78 & 1.58  \\
HD 111232   & 0.75 & $-$0.36 & 1 & 1.97  & 6.8  & HD 141937  & 1.08 &   +0.10 & 1 & 1.52 & 9.7   \\
HD 3651     & 0.76 &   +0.12 & 1 & 0.28  & 0.20 & HD 50554   & 1.09 &   +0.01 & 1 & 2.38 & 4.9   \\
HD 114783   & 0.77 &   +0.09 & 1 & 1.2   & 0.99 & BD-10 3166 & 1.10 &   +0.33 & 3 & 0.05 & 0.48  \\
OGLE-TR-113 & 0.77 &   +0.14 & 5 & 0.023 & 1.35 & HD 76700   & 1.10 &   +0.41 & 1 & 0.05 & 0.20  \\
HD 37605    & 0.80 &   +0.39 & 4 & 0.26  & 2.84 & HD 160691  & 1.10 &   +0.32 & 1 & 4.17 & 3.1   \\
HD 6434     & 0.82 & $-$0.52 & 1 & 0.15  & 0.48 & OGLE-TR-56 & 1.10 &   +0.17 & 4 & 0.023& 1.45  \\
HD 46375    & 0.82 &   +0.20 & 1 & 0.04  & 0.25 & HD 147513  & 1.11 &   +0.06 & 1 & 1.26 & 1.0   \\
OGLE-TR-111 & 0.82 &   +0.12 & 4 & 0.05  & 0.53 & HD 92788   & 1.12 &   +0.32 & 1 & 0.94 & 3.8   \\
HD 27442    & 0.83 &   +0.42 & 2 & 1.18  & 1.28 & HD 10647   & 1.14 & $-$0.03 & 1 & 2.1  & 0.91  \\
HD 190228   & 0.83 & $-$0.24 & 2 & 2.31  & 4.99 & HD 196050  & 1.15 &   +0.22 & 1 & 2.5  & 3.0   \\
HD 4208     & 0.86 & $-$0.24 & 1 & 1.67  & 0.80 & HD 209458  & 1.15 &   +0.02 & 1 & 0.05 & 0.69  \\
HD 75732    & 0.87 &   +0.33 & 1 & 5.9   & 4.05 & HD 33636   & 1.16 & $-$0.08 & 1 & 3.56 & 9.28  \\
HD 168746   & 0.88 & $-$0.08 & 1 & 0.065 & 0.23 & HD 72659   & 1.16 &   +0.03 & 1 & 3.24 & 2.55  \\
HD 145675   & 0.90 &   +0.43 & 1 & 2.8   & 4.74 & HD 121504  & 1.17 &   +0.16 & 1 & 0.32 & 0.89  \\
HD 216770   & 0.91 &   +0.26 & 1 & 0.46  & 0.65 & HD 150706  & 1.17 & $-$0.01 & 1 & 0.82 & 1.0   \\
HD 210277   & 0.92 &   +0.19 & 1 & 1.10  & 1.28 & HD 154857  & 1.17 & $-$0.23 & 4 & 1.11 & 1.80  \\
HD 83443    & 0.93 &   +0.35 & 1 & 0.04  & 0.41 & HD 2039    & 1.18 &   +0.32 & 1 & 2.19 & 4.85  \\
HD 117176   & 0.93 & $-$0.06 & 1 & 0.48  & 7.44 & HD 8574    & 1.18 &   +0.06 & 1 & 0.76 & 2.23  \\
HD 65216    & 0.94 & $-$0.12 & 1 & 1.37  & 1.21 & HD 68988   & 1.18 &   +0.36 & 1 & 0.07 & 1.90  \\
HD 102117   & 0.95 &   +0.18 & 4 & 0.15  & 0.18 & HD 52265   & 1.20 &   +0.23 & 1 & 0.49 & 1.13  \\
HD 143761   & 0.95 & $-$0.21 & 1 & 0.23  & 1.10 & HD 82943   & 1.20 &   +0.30 & 1 & 1.16 & 1.63  \\
HD 130322   & 0.96 &   +0.03 & 1 & 0.09  & 1.08 & HD 88133   & 1.20 & $-$0.34 & 4 & 0.047& 0.22  \\
HD 168443   & 0.96 &   +0.06 & 1 & 0.29  & 7.2  & HD 216437  & 1.20 &   +0.25 & 1 & 2.7  & 2.1    \\
HD 190360A  & 0.96 &   +0.24 & 1 & 4.8   & 1.33 & HD 20367   & 1.21 &   +0.17 & 1 & 1.25 & 1.07  \\
HD 108874   & 0.97 &   +0.23 & 1 & 1.07  & 1.65 & HD 40979   & 1.21 &   +0.21 & 1 & 0.81 & 3.32  \\
HD 114729   & 0.97 & $-$0.25 & 1 & 2.08  & 0.82 & HD 10697   & 1.22 &   +0.14 & 1 & 2.13 & 6.12  \\
HD 28185    & 0.98 &   +0.22 & 1 & 1.0   & 5.7 & HD 213240  & 1.22 &   +0.17 & 1 & 2.03 & 4.5   \\
HD 73256    & 0.98 &   +0.26 & 1 & 0.04  & 1.85 & HD 75289   & 1.23 &   +0.28 & 1 & 0.05 & 0.42  \\
HD 178911B  & 0.98 &   +0.27 & 1 & 0.32  & 6.29 & HD 142415  & 1.26 &   +0.21 & 1 & 1.05 & 1.62  \\
HD 1237     & 0.99 &   +0.12 & 1 & 0.49  & 3.21 & HD 74156   & 1.27 &   +0.16 & 1 & 3.4  & $>$ 6.17  \\
HD 70642    & 0.99 &   +0.18 & 1 & 3.3   & 2.0   & HD 108147  & 1.27 &   +0.20 & 1 & 0.10 & 0.41  \\
HD 186427   & 0.99 &   +0.08 & 1 & 1.72  & 1.69 & HD 142     & 1.28 &   +0.14 & 1 & 0.98 & 1.36  \\
Sun         & 1.00 &   +0.00 &   & 5.2   & 1.00 & HD 179949  & 1.28 &   +0.22 & 1 & 0.05 & 0.84  \\
HD 23079    & 1.01 & $-$0.11 & 1 & 1.65  & 2.61 & HD 9826    & 1.30 &   +0.13 & 1 & 2.53 & 3.75  \\
HD 30177    & 1.01 &   +0.39 & 1 & 3.86  & 9.17 & HD 23596   & 1.30 &   +0.31 & 1 & 2.72 & 7.19  \\
HD 106252   & 1.02 & $-$0.01 & 1 & 2.61  & 6.81 & HD 208487  & 1.30 & $-$0.06 & 4 & 0.52 & 0.43  \\
HD 217107   & 1.02 &   +0.37 & 1 & 0.07  & 1.28 & HD 17051   & 1.32 &   +0.26 & 1 & 0.93 & 2.26  \\
HD 222582   & 1.02 &   +0.05 & 1 & 1.35  & 5.11 & HD 120136  & 1.33 &   +0.23 & 1 & 0.05 & 3.87  \\
HD 177830   & 1.03 &   +0.36 & 3 & 1.0   & 1.28 & HD 216435  & 1.34 &   +0.24 & 1 & 2.7  & 1.49  \\
HD 49674    & 1.04 &   +0.33 & 1 & 0.06  & 0.12 & OGLE-TR-132& 1.34 &   +0.43 & 5 & 0.031& 1.19  \\
HD 80606    & 1.04 &   +0.32 & 1 & 0.44  & 3.41 & HD 19994   & 1.37 &   +0.24 & 1 & 1.3  & 2.0   \\
HD 187123   & 1.04 &   +0.13 & 1 & 0.04  & 0.52 & HD 169830  & 1.43 &   +0.21 & 1 & 3.6  & 4.04  \\
HD 219542B  & 1.04 &   +0.17 & 1 & 0.46  & 0.30 & HD 104985  & 1.50 & $-$0.35 & 4 & 0.78 & 6.3   \\
HD 12661    & 1.05 &   +0.36 & 1 & 0.83  & 2.30 & HD 89744   & 1.53 &   +0.22 & 1 & 0.88 & 7.20  \\
HD 16141    & 1.05 &   +0.15 & 1 & 0.35  & 0.23 & HD 38529   & 1.60 &   +0.40 & 1 & 0.13 & 0.78  \\
\hline
\end{tabular}
\medskip

Data on planet orbits and masses was obtained at http://www.obspm.fr/encycl/catalog.html; 

data on stellar metallicity and mass was obtained at the following references: 1= Santos et al. 2004; 2 = Santos et al. 2003; 

3 = Gonzalez et al. 2001; 4 = http://www.obspm.fr/encycl/catalog.html; 5 = Bouchy et al. 2004.
\end{minipage}
\end{table*}

The stellar age range of the sample thus assembled is a significant issue when one considers the timescale of the migration mechanisms, which act before the proto-planetary disc fully dissipates. We have plotted the stars of Table I in the theoretical HR diagrams of Schaerer et al. and references therein (1993) and have thus inferred stellar ages. These determinations suffer from heterogeneities in the literature database of metallicity and effective temperature (for the latter, the sources of metallicity were used), besides the uncertainties in the models themselves. Given the fact that the ages thus determined have probable uncertainties of a few billion years (Gyr), we have divided the sample of Table I in stars termed ``young", those with ages inferior to 1-2$\,$Gyr, those ``middle-aged" like the Sun, with ages within 2 and 6 Gyr, and the ``old" stars, older than 6$\,$Gyr. No trend emerged between stellar ages and the planetary orbit parameters. Actually, examples of very young stars in our sample are compatible with ages of 0.3$\,$Gyr or less (Porto de Mello \& da Silva 1997, for HD147513; K\"urster et al. 2000, for HD17051; see also Mayor et al. 2004), and it is therefore reasonable to assume that migration processes stopped long ago, producing multiple populations. However, it is also reasonable to consider, given the absence of significant observational constraints for the time being, that the migration mechanism is the same for all stars. Although there is not yet consensus on how the migration process stops, saving the planets (at least the ones that have been observed) from being swallowed by their parent stars, a correlation between SMA and metallicity could still be seen after the migration stops, for the ones that suffered a slight migration process. Thus, our model seeks to fit the observed distribution of planetary SMA as a function of stellar metallicity for the (almost) unmigrated planets. For the planets that suffered an intense migration process, it would be more difficult to distinguish between the different populations.

Fig.$\,$1 shows the metallicity versus SMA plot for the most massive planets orbiting the 72 stars with $1\pm 0.2$ solar masses. Jupiter is also included (at 5.2$\,$AU), and is assumed to be an example of a planet that suffered little or no migration (Franklin \& Soper 2003). There is a growth curve, in light of our model, representing the planets that suffered mild or no migration, and distinct populations of planets that have suffered more severe migration. Even if we constrain even more the mass range, to $1\pm 0.1$ solar masses, the growth curve is unaltered, as shown in Fig.$\,$2.

The fitting process of Eq.$\,$(\ref{eq9}), using Eq.$\,$(\ref{eq11}), produces a ZAPO curve ($n\,$=1 in Eq.$\,$(\ref{eq13})) as shown in Fig.$\,$3 where  the SMA of Jupiter and of the planets orbiting HD$\,$75732 (5.9$\,$AU) have been considered as constraints (almost unmigrated ones). We should take some value of $\alpha_{\rm a}$ between 0.5 and 1.5 and $\beta_{\rm a}$ between 1 and 2 ({\it cf.} Eq.$\,$(\ref{eq2}) and (\ref{eq3})); the constant $c$ in Eq.$\,$(\ref{eq7b}) is inversely proportional to some range of values for $Z$ where the ZAPO curve grows between the asymptotic limits established by Eq.$\,$(\ref{eq10}); $Z_0$ is the value for the point of inflection of the growing curve. 

The general behaviour of the ZAPO and temperature profiles (Eq.$\,$(\ref{eq12})) according to the parameters is the following. Assuming given values of $c$, $\beta$ and $\alpha$, and (suitably) changing the values of $\gamma$, $Z_0$ and $t$, we can fit the data with almost the same curve at the high metallicity range while the minimum $r_{\rm opt}(Z)$ varies with $\gamma$ (at the low metallicity range). However, low values of $\gamma$ are preferred in order to produce low values of temperature for low values of Z and high radius (and thus low values of $\beta$ are also preferred).  Analogous simulations, now fixing $\gamma$ and changing $c$ (so changing only the steepness of the curve), show that low values of $c$ are preferred to obtain the above behaviour of the temperature.

In order to restrict the choice of the parameters, we have done simultaneous fit of the midplane temperature profile of Nomura \& Millar (2005, their figure 2) (also D'Alessio et al. 2001, their figure 5a), assuming [Fe/H]$=0$, and of the ZAPO ($n\,$=1), having the two planets cited above as constraints, and also using the requirement that for low values of the metallicity the midplane temperature profile should be physically reasonable. For a rather good simultaneous fit we have $\alpha_{\rm a}=0.8$, $\beta_{\rm a}=1.0$, $c=4$, $t=0.028$, $Z_0 = - 0.63$ and $\gamma=1.433$ (which means that the minimum $r_{\rm opt}(Z)$ equals 150 R$_\odot$, or $\sim\,$0.7$\,$AU). Note that this distance is in agreement with the location of the puffed-up inner rim below which dust evaporation destroys the dust particles (Isella \& Natta 2005, Gorti \& Hollenbach 2004). As mentioned before, for this analysis, it is not necessary to take into account very low values of [Fe/H]. We found that, for [Fe/H]$\,\lower 3pt\hbox{$\buildrel<\over\sim$}-1$, the distribution $P(r)$ becomes very narrow and with low maximum values around values of $r$ below 1AU, such that the total probability becomes negligible.

It is worth noting that, in this formalism, $t$ mathematically defines the midplane temperature profile (Eq.$\,$(\ref{eq3})) in a given range of the disc radius, so it does not consider the temperature behaviour at the end of the disc. We believe that $t$, a small quantity, may also play the role of the dynamical parameters of the stellar disc, and thus account for ZAPO curves for different stellar masses. 

Distinct populations of migrated planets are obtained by manipulation of the relation $r(Z,n) = n\,r_{\rm opt}(Z)$. This procedure gives the ZAPO ($n\,$=1) and some of the type II migration curves ($n$$\,<\,$1) on Fig.$\,$3, for the most massive planets orbiting stars with masses around one solar mass. We can see that down to $n = 0.01$ there is no clear stratification of planet masses, as advanced in the beginning of this section.

\begin{figure}
 \includegraphics[]{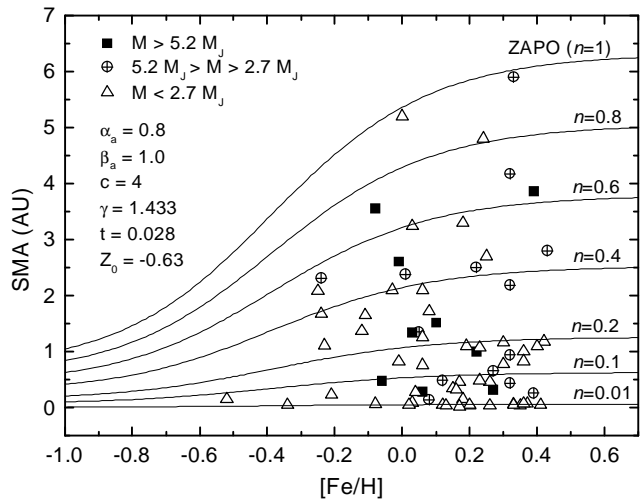}
 \caption{ZAPO curve and its evolution due to migration, $r(Z,n) = n\,r_{\rm opt}(Z)$, obtained from Eq.$\,$(\ref{eq7b}) to (\ref{eq9}), and (\ref{eq11}). The planets were displayed according to three mass ranges, where ${\rm M}_{\rm J}$ stands for one Jupiter mass.}
\end{figure}

We must also bear in mind that a mild pollution effect may have played a part in distorting the curves, particularly for the population of planets with tighter orbits, since their central stars must have engulfed more material (gas, dust and planets) in order to pull the (remaining) planets that close. Evidence that this possible offset in metallicity is lower than 0.1 dex in [Fe/H] is provided by Santos et al. (2003) and Pinsonneault et al. (2001), who show the absence of any trend of possible metallicity enhancements with the mass of stellar convective envelope. An effect of this magnitude would only displace the stars slightly to the right, without altering the global shape of the data points and the curves.

\section{Conclusions}

The sample of extrasolar planets around stars with masses similar to the Sun produces a distinct population of migrated planets. However, the ZAPO curve, which is all that we hope to model so far, is necessarily a tentative one, since we have a scarce number of planets with
larger SMA. This is probably because the Doppler shift technique requires more observational time in order to pinpoint larger orbits, and we expect that in the near future the ZAPO curve will be more easily delineated. It is also supposed, in light of the core accretion mechanism, that the larger orbits will not exceed much the current maximum, since the steep decline in dust surface density would require too much time for planets to form in the outer regions of the protoplanetary disc. The disc instability simulations, on the other hand (Mayer et al. 2002; Boss 2002b), indicate that large planets could form at even larger distances ($>\,$10$\,$AU).

The scenario of our model predicts that the metal-poorer stars have fewer giant planets not only because the dust surface density profile is affected, but also because the ones formed lie initially closer to the central stars, and have therefore been more frequently absorbed by the central stars as a consequence of the migration process. Given the statistical significance of the lack of long period planets in metal-poor stars, this is a prediction of our model that may be borne out by future observations, as the sample of detected extrasolar planets steadily increases. Also, we note that an enhanced probability of planet engulfment for the metal-poor stars would reflect in an enhanced efficiency of effectively observed planetary companions in metal-rich stars. The distribution of planet hosts would qualitatively be a steepened version of the ZAPO curves, since for metal rich stars the migration process would tend to leave behind multiple populations of survivors, but for metal-poor stars the close-in optimum formation radius would make for a more efficient engulfment of planets, leaving behind an observational void. Qualitatively, this is in very good agreement with the results of Santos et al. (2004; see their figure 7) and also explains recent statistical studies by Sozetti (2004) indicating that close-in planets are more likely to be found around metal-rich stars. Therefore, the arguments used to put forward the ZAPO model can explain the observed frequency of planetary companions with metallicity, but the argument cannot be turned around to obtain the former from the latter. Moreover, the model  provides a physical explanation for the lack of long period planets in metal-poor stars.

We have to emphasise that the model presented by this work is very simple, and that undoubtedly other factors should enter the picture in the future, as more observations become available. For instance, the quantity $P$, that we called ``probability of giant planet formation'' is likely to be related to the total density of solids, not only dust. It is a known fact that the density of solids in a disc meets a maximum where water vapour condenses into water ice. At this distance - the snowline -, the most abundant volatile becomes solid and its contribution dominates over that of rock. For a disc of solar abundance, the presence of ice exterior to the snowline increases solid surface density by a factor of four. Stevenson \& Lunine (1988) propose that diffusive redistribution of water vapour through the snowline further enhances the density of solids by 1 to 2 orders of magnitude, naturally leading to the emergence of a ``best radius" for giant planet formation. In this sense, $\sigma/T$ can be regarded as an approximation to the exact formulation $\,\sigma_{\rm dust} + \sigma_{\rm ices}\,$, with $1/T$ reproducing the process of condensation of water vapour into ice as the temperature decreases. It is interesting to note that although the discontinuous behaviour of the condensation does not emerge from this linear approximation, the location of the peak of our probability profiles (for solar and higher metallicity) reasonably agrees with the location where the density of solids peaks when ice condensation is taken into account (Kornet et al. 2004).

We acknowledge the fact that the parameter $n$, which embodies our ignorance on the migration process, will probably turn out to be quite complex. The very fitting process of the curves shown in Fig.$\,$3 suggests that the so called type II migration process is not the only one that influenced the final orbits. However, even with the approximations assumed in this work, we believe that the model, coupled with a metallicity versus SMA plot, is a simple and useful tool for probing the evolution of planetary populations.

The model predicts the optimum region for planet formation, but offers no clue as to the mechanism that triggers the process. A possible answer lies in the theory that the general radial gas density profile is probably subject to many local density enhancements, which attract solids due to pressure gradients and trigger planet formation (Haghighipour \& Boss 2003). Therefore, the density enhancement located near or at the optimum point would naturally evolve more rapidly.

Finally, it is also worth mentioning that the same model presented in this work can be tested against the formation of rocky planets, as soon as sizable samples of these become available, only that this time the temperature factor shifts to the temperature that allows the existence of dust particles.

\section*{Acknowledgments}

The authors thank the anonymous referee for very constructive criticism and suggestions,
which considerably improved the final version of this paper, and Helio J. Rocha-Pinto for the critical reading of the manuscript. GFPM acknowledges financial support by FAPERJ (grant E-26/170.687/2004), CNPq (grant 552331/01-5), FAPESP (grant 00/06769-4, USP), and by the MEGALIT/Instituto do Mil\^enio program.

\bsp

\label{lastpage}


\begin{thebibliography}{99}

\bibitem[]{} Armitage P. J., 2000, A\&A, 362, 968

\bibitem[]{} Bary J. S., Weintraub D. A., Kastner J. H, 2003, ApJ, 586, 1136

\bibitem[]{} Bodenheimer P., Pollack J. B., 1986, Icarus, 67, 391

\bibitem[]{} Bonnell I. A., Smith K. W., Davies M. B., Horne K., 2001, MNRAS, 322, 859

\bibitem[]{} Boss A. P., 2002a, ApJ, 567, L149

\bibitem[]{} Boss A. P., 2002b, ApJ, 576, 462

\bibitem[]{} Bouchy F., Pont F., Santos N.C., Melo C., Mayor M., Queloz D., Udry S., 2004, A\&A, 421, L13 

\bibitem[]{} D'Alessio P., Calvet N., Hartmann L., 2001, ApJ, 553, 321

\bibitem[]{} De Pater I., Lissauer J. J., 2001, Condensation and Growth of Solid Bodies. In Planetary Sciences pp 453-459. Cambridge University Press, Cambridge

\bibitem[]{} Franklin F. A., Soper P. R., 2003, AJ, 125, 2678

\bibitem[]{} Gilliland R. L., Brown T. M., Guhathakurta P., Sarajedini A., Milone E. F., Albrow M. D., Baliber N. R.; Bruntt H., Burrows A., Charbonneau D., Choi P., Cochran W. D., Edmonds P. D., Frandsen S., Howell J. H., Lin D. N. C., Marcy G. W., Mayor M., Naef D., Sigurdsson S., Stagg C. R., VandenBerg D. A., Vogt S. S., Williams M. D., 2000, ApJ, 545, L47

\bibitem[]{} Gonzalez G., Laws C., Tyagi S., Reddy B. E, 2001, AJ, 121, 432

\bibitem[]{} Gorti U., Hollenbach D., 2004, ApJ, 613, 424

\bibitem[]{} Haghighipour N., Boss A. P., 2003, ApJ, 583, 996

\bibitem[]{} Haisch K. E. J., Lada E. A., Lada C. J., 2001, ApJ, 553, L153

\bibitem[]{} Hartmann L., 1998, Disk Accretion. In Accretion Processes in Star Formation pp 77-101. Cambridge University Press, Cambridge

\bibitem[]{} Hurley J.R., Shara M.M, 2002, ApJ, 565, 1251

\bibitem[]{} Isella A., Natta A., 2005, 438,899

\bibitem[]{} Israelian G., Santos N. C., Mayor M., Rebolo R., 2001, Nat, 411, 163

\bibitem[]{} Jorissen A., Mayor M., Udry S., 2001, A\&A, 379, 992

\bibitem[]{} Kornet K., Rozyczka M., Stepinski T.F.,  2004, A\&A 417, 151

\bibitem[]{} K\"urster M., Endl M., Els S., Hatzes A. P., Cochran W. D., Döbereiner S., Dennerl K., 2000, A\&A, 353, L33

\bibitem[]{} Laughlin G., 2000, ApJ, 545, 1064

\bibitem[]{} Lineweaver C. H., 2001, Icarus, 151, 307

\bibitem[]{} Matsuyama I., Johnstone D., Murray N., 2003, ApJ, 585, L143

\bibitem[]{} Mayer L., Quinn T., Wadsley J., Stadel J., 2002, Sci, 298, 1756

\bibitem[]{} Mayor M., Udry S., Naef D., Pepe F., Queloz D., Santos N. C., Burnet M., 2004, A\&A, 415, 391

\bibitem[]{} Mayor M., Queloz D., 1995, Nat 378, 355

\bibitem[]{} Mizuno H., 1980, Prog Theor Phys, Vol. 64, 544

\bibitem[]{} Murray N., Chaboyer B., 2002, ApJ, 566, 442

\bibitem[]{} Nomura H., Millar T. J., 2005,  A\&A,  438, 923

\bibitem[]{} Pinsonneault M. H., DePoy D. L., Coffee M., 2001, ApJ, 556, L59

\bibitem[]{} Pollack J. B., Hubickyj O., Bodenheimer P., Lissauer J. J., Podolak M., Greenzweig Y., 1996, Icarus, 124, 62

\bibitem[]{} Porto de Mello G. F., da Silva L., 1997, ApJ, 476, L89

\bibitem[]{} Pringle J. E., 1981, ARA\&A, 19, 137

\bibitem[]{} Rice W. K. M., Armitage P. J., 2003, ApJ, 598, L55

\bibitem[]{} Santos N. C., Israelian G., Mayor M., 2004, A\&A, 415, 1153

\bibitem[]{} Santos N. C., Israelian G., Mayor M., Rebolo R., 2003, A\&A, 398, 363

\bibitem[]{} Schaerer D., Charbonnel C., Meynet G., Maeder A., Schaller G., 1993, A\&AS, 98, 523

\bibitem[]{} Schneider J., 2005, Extrasolar Planets Encyclopedia: www.obspm.fr/planets

\bibitem[]{} Sigurdsson S., 1992, ApJ, 399, 95

\bibitem[]{} Sozetti, A., 2004, MNRAS, 354, 1194

\bibitem[]{} Stevenson D.J., Lunine J.I., 1988, Icarus, 75, 146

\bibitem[]{} Takeuchi T., Lin D. N. C., 2002, ApJ, 581, 1344

\bibitem[]{} Udry S., Mayor M., Clausen J. V., Freyhammer L. M., Helt B. E., Lovis C., Naef D., Olsen E. H., Pepe F., Queloz D., Santos N. C., 2003, A\&A, 407, 679

\bibitem[]{} Weldrake D. T. F., Sackett P. D., Bridges T. J., Freeman K. C., 2005, ApJ, 620, 1043

\bibitem[]{} Wetherill G. W., 1996, Icarus, 119, 219

\bibitem[]{} Youdin A. N., Shu F. H., 2002, ApJ, 580, 494

\bibitem[]{} Zucker S., Mazeh T., 2002, ApJ, 568, L113



\end{thebibliography}
\end{document}